\documentclass[prd,aps,secnumarabic,showpacs]{revtex4}
\usepackage{bm,latexsym,amsmath,amsfonts,axodraw}
\usepackage{hyperref,url}
\usepackage[dvips]{graphicx,color}
\def\ds{\displaystyle}
\begin{document}
\title{Generalized Hidden $\mathcal{Z}_2$ Symmetry of Neutrino Mixing}
\author{Duane A. Dicus$^{1,}$\footnote{Electronic address: dicus@physics.utexas.edu}, Shao-Feng Ge$^{1,2,}$\footnote{Electronic address: gesf02@gmail.com}, and Wayne W. Repko$^{3,}$\footnote{Electronic address: repko@pa.msu.edu}} \affiliation{$^1$Physics Department, University of Texas, Austin, TX 78712  \\ $^2$Center for High Energy Physics, Tsinghua University, Beijing 100084, China  \\ $^3$Department of Physics and Astronomy, Michigan State University, East Lansing MI 48824}
\date{\today}
\begin{abstract}
We explore the consequences of the neutrino mass matrix having a hidden $\mathcal{Z}_2$ symmetry and one zero eigenvalue. When implemented, these two conditions give relations among the mixing angles.  In addition, fitting these relations to the existing oscillation data allows limits to be placed on the parameter of the symmetry.
\end{abstract}
\maketitle
\section{Introduction}

\label{sec:Intro}

Neutrino physics can anticipate an era of higher precision measurements
with the upcoming generation of neutrino experiments. In the past, measurements have shown that the mixing pattern of lepton sector is quite different from that of quark sector. In the Pontecorvo-Maki-Nakagawa-Sakata (PMNS) parameterization \cite{PMNS} for lepton mixing there are two large mixing angles. The atmospheric mixing angle $\theta_a \equiv \theta_{23}$ is almost maximal while the solar mixing angle $\theta_s \equiv \theta_{12}$ is also large and the reactor mixing angle $\theta_x \equiv \theta_{31}$ nearly vanishes. The recent results are summarized in Table\,\ref{tab:data}. We can see that the uncertainties in mixing angles are not particularly small. In most measurements there is roughly a $3^\circ$ deviation at $1 \sigma$ confidence level.
\begin{table*}[h]
\begin{center}

{\small
\begin{tabular}{c|c|c|c|c|c}
\hline
& & & & & \\

& $\Delta_s\,(10^{-5}{\rm eV}^2)$

& $\Delta_a\,(10^{-3}{\rm eV}^2)$

& $\sin^2\theta_s\,(\theta_s)$

& $\sin^2\theta_a\,(\theta_a)$

& $\sin^2\theta_x\,(\theta_x)$

\\

& & & & & \\[-2mm]

\hline

& & & & & \\[-2mm]

{Central Value} & $7.67$ & $2.39$ & $0.312$~($34.0^\circ$) &

$0.466$~($43.0^\circ$) & $0.016$~($7.3^\circ$)

\\[1mm]

\hline

& & & & & \\[-2mm]

$1\,\sigma$ Range & $7.48-7.83$ & $2.31-2.50$ &

$0.294 - 0.331$ & $0.408 - 0.539$ & $0.006 - 0.026$

\\[0.6mm]

& & & ($32.8-35.1^\circ$) & ($39.7-47.2^\circ$) &  ($4.4-9.3^\circ$)

\\[1mm]

\hline

\end{tabular}

\caption{The global $3\nu$ fit\,\cite{Fogli-08} for the neutrino mass-squared differences and mixing angles including the available data from solar, atmospheric, reactor (KamLAND and Chooz) and accelerator (K2K and MINOS) experiments\,\cite{nu2008}.}

\label{tab:data}}
\end{center}
\end{table*}
The mixing matrix which incorporates these angles and diagonalizes the mass matrix $M_{\nu}$ via $U^TM_{\nu}U\,=\,M^{{\rm diag}}_\nu$ is given by

\begin{equation}
U_\nu=\begin{pmatrix}
        c_sc_x & -s_sc_x & -s_xe^{i\delta_D} \\
        s_sc_a-c_ss_as_xe^{-i\delta_D} & c_sc_a+s_ss_as_xe^{-i\delta_D} & -s_ac_x \\
        s_ss_a+c_sc_as_xe^{-i\delta_D} & c_ss_a-s_sc_as_xe^{-i\delta_D} & c_ac_x \end{pmatrix} \label{Utot}
\end{equation}

with $(s_{\alpha},c_{\alpha})\,\equiv\,(\sin\theta_{\alpha},\cos\theta_{\alpha})$ for $\alpha\,=\,s,a,x$.  $\delta_D$ is the Dirac phase and we have neglected Majorana phases. From Table I we see that a good first approximation is to take $\theta_x\,=\,0,\,\, \theta_a\,=\,45^{\circ}$ which gives
\begin{equation}
U_\nu(\theta_s)=\begin{pmatrix}
 \cos \theta_s & - \sin \theta_s & 0 \\
 \sqrt{\frac 1 2} \sin \theta_s & \sqrt{\frac 1 2} \cos \theta_s & - \sqrt{\frac 1 2} \\
 \sqrt{\frac 1 2} \sin \theta_s & \sqrt{\frac 1 2} \cos \theta_s &   \sqrt{\frac 1 2}\end{pmatrix}\equiv\begin{pmatrix}
                                     v_1 & v_2 & v_3\end{pmatrix}\,.
\label{eq:Unus}
\end{equation}
Using this as a starting point, we wish to investigate whether there is an underlying symmetry $G$ of the neutrino mass matrix. This matrix must satisfy $[G,M_\nu]=0$, or $G^TM_\nu G=M_\nu$. Given a $G$, the transformation $GU_\nu$ also diagonalizes $M_\nu$. But $U_\nu$ is unique except for phases. This can be seen by supposing that $d_\nu$ is a unitary matrix such that $U_\nu d_\nu$ also diagonalizes $M_\nu$. For this to be true, $d_\nu$ must satisfy $d^T_\nu M^{{\rm diag}}_\nu d_\nu=M^{{\rm diag}}_\nu$ and this implies that $d_\nu^2=1$. Since $GU_\nu$ diagonalizes $M_\nu$, it must have the form $GU_\nu=U_\nu d_\nu$, where $d_\nu$ has $1$ or $-1$ diagonal elements. Thus 
\begin{equation}
U^\dagger_\nu G U_\nu=d_\nu\equiv
\begin{pmatrix}
       d_1 \\
     & d_2 \\
     & & d_3
\end{pmatrix}\qquad \Leftrightarrow \qquad
G=U_\nu d_\nu U^\dagger_\nu \,.
\label{eq:UGU}
\end{equation}
\noindent
There are only eight possible combinations for the elements of $d_\nu$. Two of these are the unit matrix and its negative, both of which define $G$ as a multiple of the identity. Of the remaining six, three have two entries of $+1$ and one of $-1$, while the other three have two entries of $-1$ and one of $+1$. These diagonal matrices differ by an overall minus sign, so only one of the two types is independent. If we choose the three with one entry of $1$ and two entries of $-1$ ($\det(G)=1$), then it is easy to see that multiplying any pair of these diagonal matrices will result in the remaining matrix. Hence, in reality, only two of these matrices are independent and both represent a ${\cal Z}_2$ symmetry. Since the independent $G$'s commute, the horizontal symmetry of lepton mixing is $\mathcal Z_2 \times \mathcal Z_2$ if neutrinos are Majorana fermions \cite{Lam:2006wm,Lam:2008sh,Grimus:2009pg,GHY}.

A representation of $G$ can be obtained using
\begin{equation}
G=d_1 v_1 v^\dagger_1 + d_2 v_2 v^\dagger_2 + d_3 v_3 v^\dagger_3\,.
\label{eq:G1}
\end{equation}
Since the eigenvalue $1$ can occur in three places, there are three symmetry matrices $G$
$$
G_1=\left(\begin{array}{ccc}
    c_s^2-s_s^2       & \sqrt{2}s_sc_s   & \sqrt{2}s_sc_s \\
    \sqrt{2}s_sc_s    & -c_s^2           & s_s^2          \\
    \sqrt{2}s_sc_s    & s_s^2            & -c_s^2
    \end{array}\right)\quad G_2=\left(\begin{array}{ccc}
    -(c_s^2-s_s^2)     & -\sqrt{2}s_sc_s  & -\sqrt{2}s_sc_s \\
    -\sqrt{2}s_sc_s    & -s_s^2           & c_s^2           \\
    -\sqrt{2}s_sc_s    & c_s^2            & -s_s^2
    \end{array}\right)\,,
$$
and
$$
G_3=\left(\begin{array}{ccc}
    -1  & 0  &  0   \\
    0   & 0  &  -1  \\
    0   & -1 & 0
   \end{array}\right)\,,
$$
where the subscript $i$ on $G_i$ denotes the component of  $d^{(i)}_\nu$ that is $+1$. $G_3$ gives $\mu-\tau$ symmetry~\cite{mu-tau} while $G_1$ is symmetric and commutes with $G_3$. For simplicity we can parameterize the solar mixing angle $\theta_s$ as
\begin{equation}
\cos \theta_s\equiv\frac {- k} {\sqrt{k^2 + 2}},
\qquad \sin \theta_s\equiv\frac {\sqrt 2}{\sqrt{k^2 + 2}}.
\label{eq:solar-k}
\end{equation}
Then the mixing matrix (\ref{eq:Unus}) takes the form \cite{GHY}
\begin{equation}
U_\nu(k)=\begin{pmatrix}
\frac {-k}{\sqrt{2 + k^2}} & \frac {- \sqrt 2}{\sqrt{2 + k^2}} & 0 \\
\frac 1 {\sqrt{2 + k^2}} & \frac {- k} {\sqrt{2 (2 + k^2)}} & - \frac 1 {\sqrt 2} \\
\frac 1 {\sqrt{2 + k^2}} & \frac {- k} {\sqrt{2 (2 + k^2)}} &   \frac 1 {\sqrt 2}
\end{pmatrix}
\equiv
U_k.
\label{eq:Uk}
\end{equation}
Consequently, the symmetry transformation matrix $G_1(\theta_s)$  can be reexpressed in terms of $k$
\begin{equation}
G_1(k)=\frac{1}{2+k^2}
\begin{pmatrix}
    2-k^2 & 2k & 2k  \\
       2k & k^2 & -2  \\
       2k & -2 & k^2

\end{pmatrix}.
\label{G1}
\end{equation}

Although we can ``derive" a generalized form of $G_1$ symmetry transformation matrix (\ref{G1}) given the mixing matrix (\ref{eq:Uk}), this relationship cannot be reversed. The mixing matrix $U_\nu$ can not be uniquely determined solely by $G_1$ due to the fact that $G_1$ has degenerate eigenvalues.

Invariance under $G_3$ requires $\theta_x=0^{\circ}$, $\theta_a\,=\,45^{\circ}$, but invariance under $G_1$ does not, so below we assume that the neutrino mass matrix is invariant under $G_1$, not only in the approximation $\theta_x\,=\,0^{\circ},\,\,\theta_a\,=\,45^{\circ}$, but for general values of all the mixing angles.
In the next section we use this assumption in the form of Eq.(\ref{G1}) with general values of $k$ to derive relations among the mixing angles.
In Sec.\,3 we compare our results with the experimental values and in Sec.\,4 we summarize.

\section{Invariance under the $\mathcal{Z}_2$ symmetry $G_1$} \label{sec:concrete}

In this section we show explicitly the consequences of generalized $G_1$ symmetry. Only two mass square differences have been measured and the neutrino's mass scale has not been determined by experiments. It is possible that one of the mass eigenvalues vanishes. This is also theoretically motivated by minimal seesaw model \cite{MinimalSeesaw}. We will explore the joint consequences of one vanishing mass eigenvalue and $G_1$ invariance. For simplicity we will postpone discussion of $CP$ phases to a later article \cite{CPGDR}.

\subsection{Constraints on Mass Matrix Elements}

If the neutrinos are Majorana fermions, their mass matrix must be symmetric. We will consider the case that there are three generations of light neutrinos. Then, the most general form of the neutrino mass matrix can be parameterized as

\begin{equation}
M_{\nu}=\begin{pmatrix}
          A & B_1 & B_2  \\
        B_1 & C_1 & D  \\
        B_2 & D & C_2
\end{pmatrix}\,,\label{M}
\end{equation}
which has six independent matrix elements. We assume $M_{\nu}$ is invariant under the $G_1$ symmetry transformation,
\begin{equation}
G_1^T M_{\nu} G_1=M_{\nu}\,.
\label{GMG}
\end{equation}
With the help of (\ref{G1}) and (\ref{M}), Eq.\,(\ref{GMG}) gives two conditions on the neutrino mass matrix elements of (\ref{M}) \cite{DGR} ,
\begin{subequations}
\begin{eqnarray}
\frac{B_1+B_2}{C_1+C_2+2D-2A}\,&=&\,\frac{k}{k^2-2}\,,  \label{BCDAk1}  \\
\frac{B_1-B_2}{C_1-C_2}\,&=&\,\frac{1}{k}\,.  \label{BCDAk2}
\end{eqnarray}
\label{BCDAk}
\end{subequations}

\subsection{Eigenvalues and eigenstates}\label{sec:eigen}

If there is a vanishing mass eigenvalue $m_i = 0$ the corresponding mass eigenstate, which can be denoted as $v \equiv (\alpha, \beta, \gamma)^T$, must satisfy

\begin{equation}
\begin{pmatrix}
        A & B_1 & B_2  \\
      B_1 & C_1 & D  \\
      B_2 & D & C_2
\end{pmatrix}
\begin{pmatrix}
      \alpha \\
      \beta  \\
      \gamma
\end{pmatrix}=0\,.\label{m00}
\end{equation}
If we assume $\alpha\,\ne\,0$ then we get three equations

\begin{subequations}
\begin{eqnarray}
A\,&=&\,-\rho\,B_1-\sigma\,B_2\,,  \label{ABB}  \\
B_1\,&=&\,-\rho\,C_1-\sigma\,D\,,  \label{BCD}  \\
B_2\,&=&\,-\rho\,D-\sigma\,C_2\,,  \label{BDC}
\end{eqnarray}\label{eq:0mass}
\end{subequations}
where $\rho \equiv \beta/\alpha,\,\sigma \equiv \gamma/\alpha$. Thus we have two sets of conditions, (\ref{BCDAk}) from $G_1$ invariance, and (\ref{eq:0mass}) from the vanishing mass eigenvalue.

From the relations (\ref{eq:0mass}) we can express the matrix element $A$ in terms of $C_1$, $C_2$ and $D$
\begin{equation}
A=\rho^2 C_1+ \sigma^2 C_2+ 2 \rho \sigma \,D\,.\label{AA}
\end{equation}
Now let us use these in the $\mathcal{Z}_2$ relations.  Eq.\,(\ref{BCDAk1}) and Eq.\,(\ref{BCDAk2}) give
\begin{subequations}
\begin{eqnarray}
(\sigma\,k+1)C_2-(\rho\,k+1)C_1+k(\rho-\sigma)D=0\,,  \label{C2C1D}  \\
(\rho\,k+1)(2\rho-k)C_1+(\sigma\,k+1)(2\sigma-k)C_2+[(2-k^2)(\rho+\sigma)-2k(1-2\rho\sigma)] D=0\,.  \label{C1C2D}
\end{eqnarray}
\end{subequations}
The above two relations can be reexpressed in terms of only two matrix elements, $D$ and $C_2$ or $C_1$ respectively
\begin{subequations}
\begin{eqnarray}
(\rho+\sigma-k)[(\sigma\,k+1)C_2+(\rho\,k+1)D]\,&=&\,0\,,\label{C2D}  \\
(\rho+\sigma-k)[(\rho\,k+1)C_1+(\sigma\,k+1)D]\,&=&\,0\,. \label{C1D}
\end{eqnarray}
\end{subequations}
The mass eigenvalues that are nonzero are given by
\begin{equation}\label{Mass}
m_{\pm}\,=\,\frac{1}{2}\left[A+C_1+C_2\pm\sqrt{(A+C_1+C_2)^2+4(\rho^2+\sigma^2+1)\, (D^2-C_1C_2)}\right]\,,
\end{equation}
where we have used (\ref{BCD}) and (\ref{BDC}) and (\ref{AA}).  From (\ref{C2D}) and (\ref{C1D}) it is obvious that one possible solution to the equations for $C_1,C_2,D$ is
\begin{subequations}
\begin{eqnarray}
C_1\,&=&\,-\frac{\sigma\,k+1}{\rho\,k+1}D\,,\label{C1Dx} \\
C_2\,&=&\,-\frac{\rho\,k+1}{\sigma\,k+1}D\,. \label{C2Dx}
\end{eqnarray}
\end{subequations}
This makes $D^2=C_1C_2$ and consequently $m_{-}$ given above would also be zero. Since the experimental data shows that two mass square differences between the three neutrino mass eigenvalues are nonzero, we need at least two masses nonzero in order to have two oscillation lengths.

A second solution of (\ref{C2D}), (\ref{C1D}) is $\rho\,=\,\sigma\,=\,-1/k$ but then the three relations in (\ref{eq:0mass}) simply reproduce the conditions (\ref{BCDAk}). So the conclusion is that we must have
\begin{equation}
\rho\,=\,k-\sigma\,\label{rsk}
\end{equation}
and the conditions (\ref{C2C1D}) or (\ref{C1C2D}) reduce to an equation for $\sigma$
\begin{equation}\label{SCCD}
\sigma\,=\,\frac{(1+k^2)C_1-C_2-k^2D}{k(C_1+C_2-2D)}\,.
\end{equation}
This relation represents the constraint from $G_1$ invariance which was originally expressed as (\ref{BCDAk})
where there were two independent relations. Using (\ref{rsk}), these two relations are satisfied simultaneously and reduce to a single constraint (\ref{SCCD}).

The condition (\ref{rsk}) can also be substituted into (\ref{ABB}), (\ref{BCD}), and (\ref{BDC}) to give
\begin{subequations}
\begin{eqnarray}
\sigma\,&=&\,\frac{A+kB_1}{B_1-B_2}\,, \label{ABBk} \\
\sigma\,&=&\,\frac{B_1+kC_1}{C_1-D}\,, \label{BCDk}  \\
\sigma\,&=&\,\frac{B_2+kD}{D-C_2}\,,  \label{BDCk}
\end{eqnarray}
\end{subequations}
respectively. These three relations are a manifestation of vanishing mass eigenvalue. We can set these equations for $\sigma$ equal to get relations among the matrix elements $A,\ldots,D$ in terms of the parameter $k$. Not all of these equations are independent but two different relations are possible:
\begin{subequations}
\begin{eqnarray}
(A + k B_1) (C_1 - D)- (B_1 + k C_1)(B_1 - B_2) & = & 0\,,\label{eq:krel-1}\\
(A + k B_1) (D - C_2)- (B_2 + k D) (B_1 - B_2) & = & 0\,.\label{eq:krel-2}
\end{eqnarray}\label{eq:krel}
\end{subequations}

In the next subsection we will write $A,\ldots,D$ in terms of the mixing angles and thereby get two relations among the mixing angles, again involving $k$.

\subsection{Reconstruction of Neutrino Mass Matrix}\label{sec:restrictions}

Using $U_{\nu}$ from Eq.\,(\ref{Utot}) in $M_{\nu}\,=\,U^{*}M^{{\rm diag}}_{\nu}U^{\dagger}$ and comparing with (\ref{M}) we get \cite{BDHL}
\begin{subequations}
\begin{eqnarray}
A & = & c_x^2 c_s^2 m_1 + c_x^2 s_s^2 m_2 + s_x^2 m_3 \label{VA}\\
B_1 & = & c_x [s_s c_s c_a - s_x s_a c_s^2] m_1 - c_x [s_s c_s c_a + s_x s_a s_s^2] m_2 + c_x s_x s_a m_3 \label{VB1}\\
B_2 & = & c_x [s_s c_s s_a + s_x c_a c_s^2] m_1 - c_x [s_s c_s s_a - s_x c_a s_s^2] m_2 - s_x c_x c_a m_3 \label{VB2}\\
C_1 & = & (s_s c_a - s_x c_s s_a)^2 m_1 + (c_s c_a + s_x s_s s_a)^2 m_2 + c_x^2 s_a^2 m_3 \label{VC1} \\
C_2 & = & (s_s s_a + s_x c_s c_a)^2 m_1 + (c_s s_a - s_x s_s c_a)^2 m_2 + c_x^2 c_a^2 m_3 \label{VC2} \\
D & = & (s_s s_a + s_x c_s c_a)(s_s c_a - s_x c_s s_a) m_1 + (c_s s_a - s_x s_s c_a)(c_s c_a + s_x s_s s_a) m_2 - c_x^2 s_a c_a m_3 \label{VD}
\end{eqnarray}\label{eq:reconstruction}
\end{subequations}
where, as mentioned above, we have deferred consideration of $CP$ violation to a later article. The mass eigenvalues can be further parameterized in terms of experimentally measured mass square differences: $m_1 = m_0$, $m_2 = m_0 \sqrt{1+r}$ and $m_3 = 0$ for inverted mass hierarchy and $m_1 = 0$, $m_2 = m_0 \sqrt r$ and $m_3 = m_0$ for normal mass hierarchy where $m_0 \equiv \sqrt{\Delta_a}$ and $r\equiv\Delta_s/\Delta_a$ which is positive.

\subsection{Correlations between Mixing Angles}\label{sec:inverted}

To get relations between mixing angles we can substitute (\ref{eq:reconstruction}) into (\ref{eq:krel}) which gives
\begin{subequations}
\begin{eqnarray}
- c_a c_x \left[ c_x (c_a - s_a) + k s_x \right] m_1 m_2 & = & 0\,,\\
s_a c_x \left[ c_x (c_a - s_a) + k s_x \right] m_1 m_2 & = & 0\,.
\end{eqnarray}
\end{subequations}
where we have assumed the mass hierarchy is inverted with vanishing $m_3$ and nonzero $m_1$ and $m_2$, while $c_a$, $s_a$ and $c_x$ are also nonzero. The only possible solution is
\begin{equation}\label{r1}
k = c_x \frac {s_a - c_a}{s_x} \approx \frac {\sqrt 2 \delta_a}{\delta_x}\,,
\end{equation}
\noindent where the last factor comes from  expanding the two mixing angles $\theta_a$ and $\theta_x$ around the approximations  $45^{\circ}$ and $0^{\circ}$,
\begin{equation}
\theta_a\equiv\frac \pi 4 + \delta_a\,,\qquad\theta_x\equiv\delta_x\,.
\label{eq:expansion}
\end{equation}
With this solution for $k$ and the reconstructed mass matrix elements (\ref{eq:reconstruction}) substituted back into (\ref{ABBk}) we find
\begin{equation}\label{r2}
\frac{1}{\sigma}\,=\,\frac{c_a-s_a}{k\,c_a}=- \frac {s_x}{c_x c_a} \approx - \sqrt 2 \delta_x\,.
\end{equation}
\noindent Since $\delta_x$ is quite small, $\sigma$ should be very large according to (\ref{r2}).

\begin{figure}[ht]\centering
\includegraphics[width=3in]{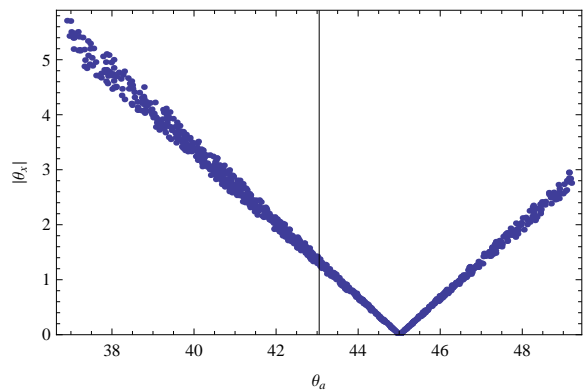}
\caption{Prediction of $\theta_x$ in terms of $\theta_a$ and $\theta_s$ at the 90\% C.L. The vertical solid line denotes the experimentally measured central value of
the atmospheric mixing angle $\theta_a$.}\label{fig:thetax}
\end{figure}

We still have the condition from $\mathcal{Z}_2$, Eq.\,(\ref{SCCD}). Together with (\ref{r2}) and (\ref{eq:reconstruction}) as well as (\ref{r1}) it gives
\begin{equation}
\tan 2 \theta_s=\frac {2 (c^2_a - s^2_a)s_x}{c^2_x - (2 + 2s^2_x) c_a s_a} = \frac {2 \left( \frac{\ds c_a + s_a}{\ds c_a - s_a} s_x \right)}
{1 - \left( \frac {\ds c_a + s_a}{\ds c_a - s_a} s_x \right)^2} =
\frac {2 \left( - \frac {\ds c_a - s_a}{\ds c_a + s_a} \frac{\ds 1} {\ds s_x} \right)}{1 - \left( - \frac {\ds c_a - s_a}{\ds c_a + s_a} \frac{\ds 1} {\ds s_x} \right)^2}\,.
\end{equation}
There are two possible solutions

\begin{equation}\label{r3p}
\tan\theta_s\,=\,- \frac {c_a - s_a}{c_a + s_a} \frac{1}{s_x} = \frac{k}{c_x(s_a+c_a)}\approx\frac {\delta_a}{\delta_x}\,,
\end{equation}
or
\begin{equation}\label{r3a}
\tan\theta_s\,=\,\frac {c_a + s_a}{c_a - s_a}s_x = - \frac{c_x(c_a+s_a)}{k} \approx - \frac {\delta_x}{\delta_a}\,.
\end{equation}
These relations between mixing angles can be used to predict the not well measured $\theta_x$ in terms of the solar and atmospheric mixing angles.
For example (\ref{r3a}) gives
\begin{eqnarray}
s_x=\frac {c_a - s_a}{s_a + c_a}\frac {s_s}{c_s}\qquad \Rightarrow \qquad \delta_x \approx - \tan \theta_s \delta_a
\label{eq:sx-b}
\end{eqnarray}
Since $\theta_x$ is the focus of next generation of neutrino experiments, we use (\ref{eq:sx-b}) to estimate its value. The scatter plot is shown in Fig. \ref{fig:thetax}. A scatter plot based on (\ref{r3p}) would look similar with a steeper slope for the points.

Another way of expressing the results is to write all of the mixing angles in terms of the parameters $\sigma$ and $k$. Using $z=1/\sigma$, this gives
\begin{subequations}
\begin{eqnarray}
\sin^2 \theta_a & = & \frac{(1-kz)^2}{k^2z^2-2kz+2}\label{s2a} \\
\sin^2 \theta_x & = & \frac{z^2}{(k^2+1)z^2-2kz+2}\label{s2x} \\
{\rm and}\,\,\,\,{\rm either}\,\,\,\,\,\,\,\,\,\,\,\,\,\,\,\,\,\,\,\,\,\,\,\,\,\,\,\,\,\,\,\,\,\,\,\,\,\,\,&&  \nonumber \\
\sin^2 \theta_s & = & \frac{(2-kz)^2}{k^4z^2-2k^3z+2k^2(z^2+1)-4kz+4} \label{s2s2}\\
{\rm or}\,\,\,\,\,\,\,\,\,\,\,\,\,\,\,\,\,\,\,\,\,\,\,\,\,\,\,\,\,\,\,\,\,\,\,\,\,\,\,\,\,\,\,\,&&  \nonumber \\
\sin^2\theta_s\,&=&\,k^2\frac{(k^2+1)z^2-2kz+2}{k^4z^2-2k^3z+2k^2(z^2+1)-4kz+4}\,,\label{s2s}
\end{eqnarray}\label{eq:s2asx-IH}
\end{subequations}
Note that these equations are all unchanged under $k,z\,\longrightarrow\,-k,-z$ so only the absolute value of $k$ can be determined.

\section{Fit to existing data}\label{sec:fit}

The solutions (\ref{r3p}) or (\ref{r3a}), which give (\ref{s2s2}) or (\ref{s2s}), are identical in the following sense - oscillation experiments measure $\sin^22\theta$ and thus can't distinguish between $\theta$ and $\pi/2-\theta$.  Further, $\tan(\pi/2-\theta)\,=\,1/\tan\theta$, so a fit with (\ref{r3p}) and $\theta_s$ assumed greater than $\pi/4$ is identical to one with (\ref{r3a}) and $\theta_s$ assumed less than $\pi/4$.
Having noted this we will proceed to fit both (\ref{s2s2}) and (\ref{s2s}) with $\theta_s\,<\,\pi/4$.

\begin{figure}[h]\centering
\includegraphics[width=2.75in]{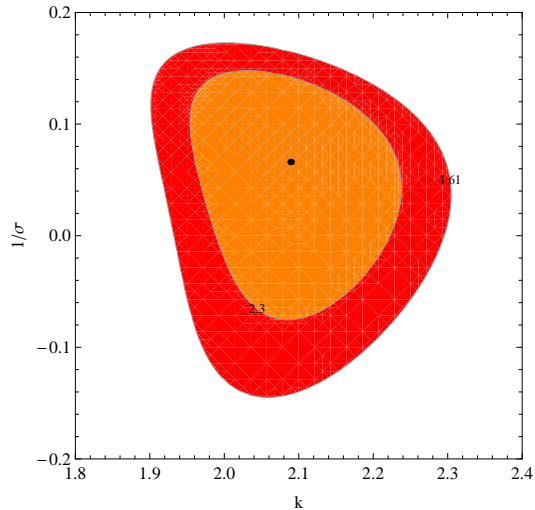}
\caption{The 68.3\% and 90.0\% confidence contours for the fit using Eqs.\,(\ref{s2a}), (\ref{s2x}), and (\ref{s2s2}) are shown in red and orange, respectively. The (black) dot indicates the $\chi^2$ minimum.} \label{abc}
\end{figure}

Using Eqs.\,(\ref{s2a}), (\ref{s2x}), and (\ref{s2s2}), the fit to the data from Ref.\,\cite{Fogli-08} gives $\chi^2_{\rm min}=2.10$, $|k|_{\rm min}=2.09$ and $z_{\rm}=0.066$. At the minimum values of $|k|$ and $z$, $\sin^2(\theta_a)=0.426\,(\theta_a=40.7^\circ)$, $\sin^2(\theta_x)=0.0025\, (\theta_x=2.87^\circ)$ and $\sin^2(\theta_s)=0.313\,(\theta_s=34.0^\circ)$. The $68.3\%$ and $90\%$ confidence contours are shown in Fig.\,\ref{abc}.

\begin{figure}[h]\centering
\includegraphics[width=2.3in]{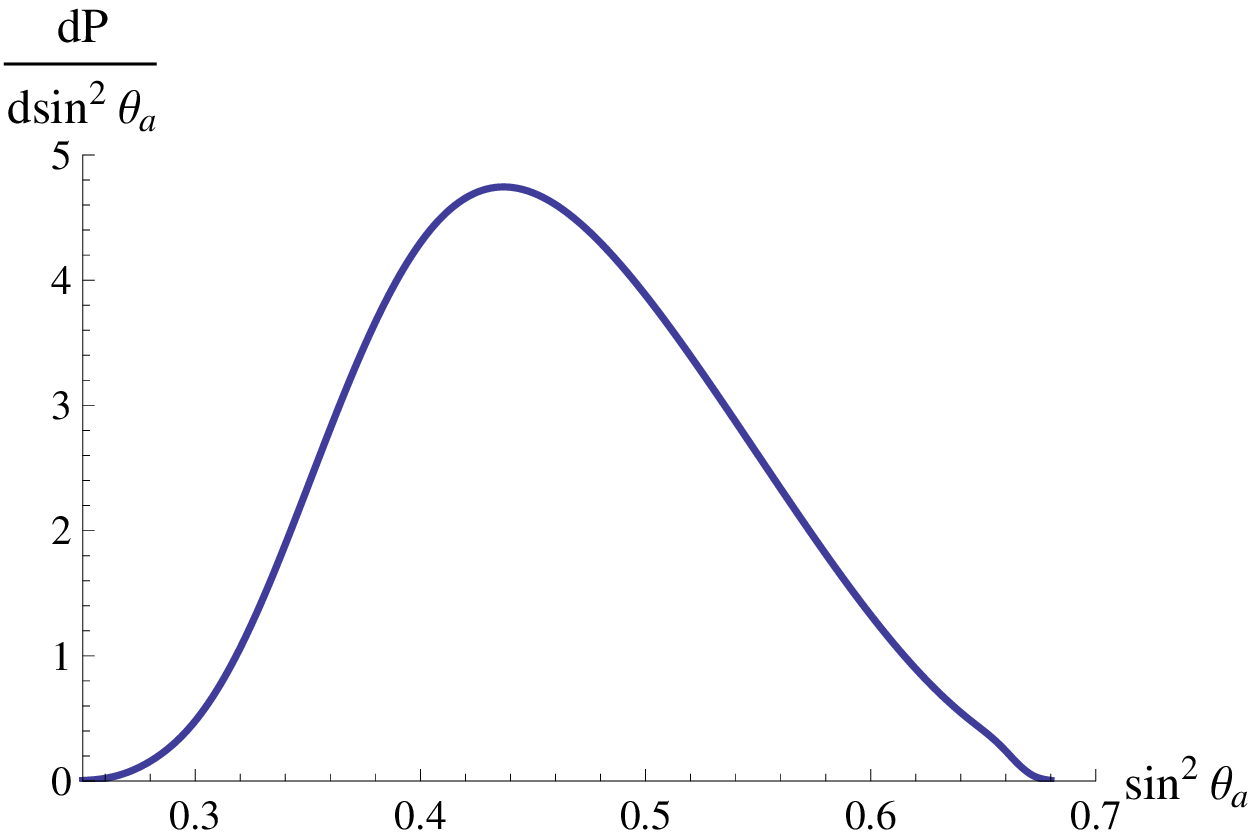}
\hfill\includegraphics[width=2.3in]{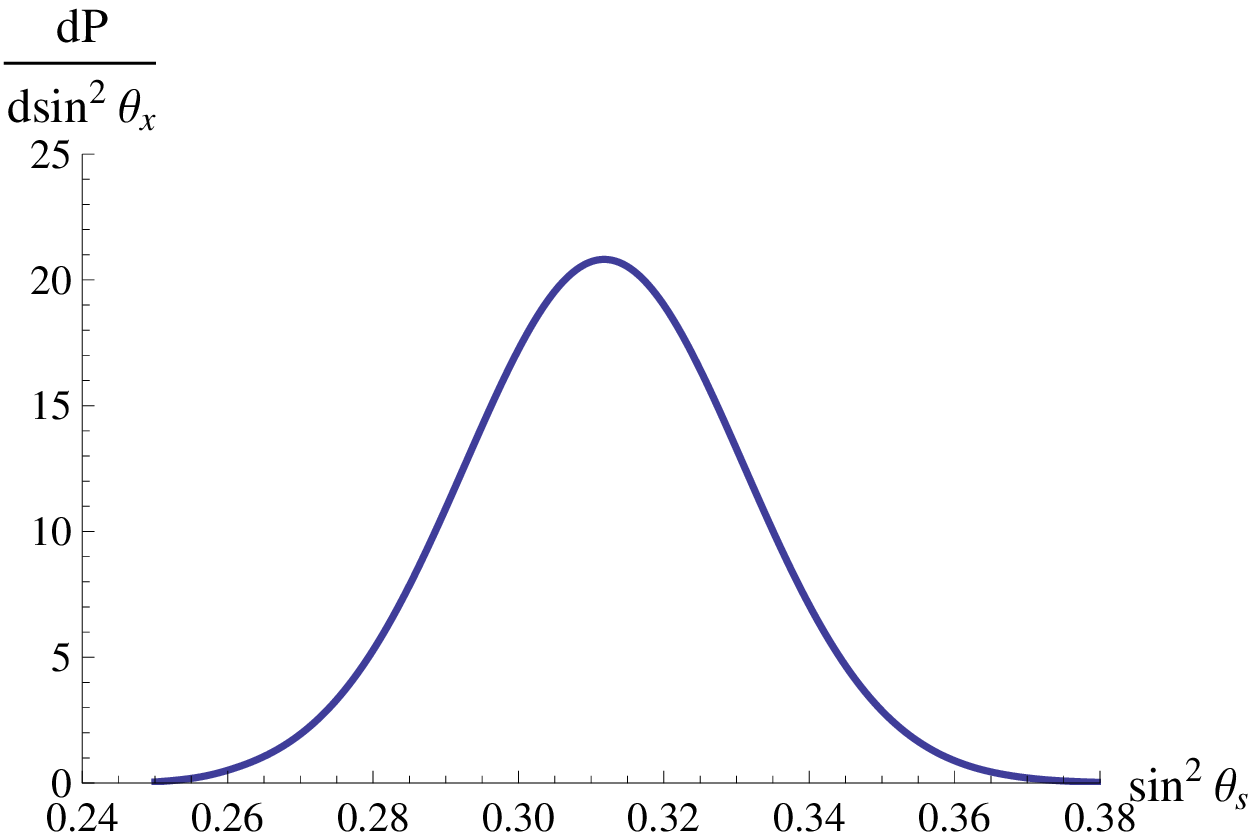}\hfill\vspace{6 pt} 
\hfill\includegraphics[width=2.3in]{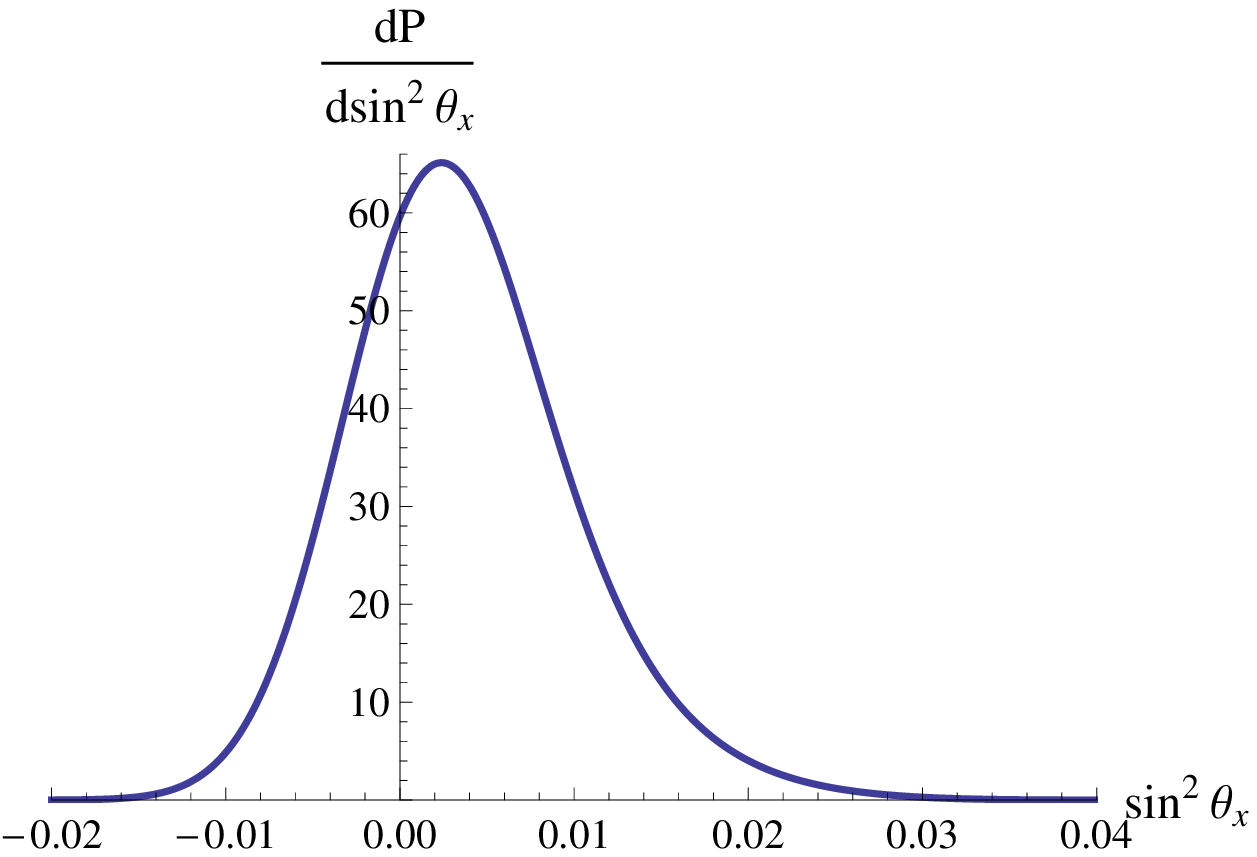}\hfill
\caption{The distributions of the $\sin^2\theta_i$ obtained using Eqs.\,(\ref{s2a}), (\ref{s2x}), and (\ref{s2s2}) are shown.}\label{distna}
\end{figure}

The distributions of this set of mixing angles are obtained from the likelihood distribution 
\begin{equation}
Ae^{-(\chi^2(k,z)-\chi^2_{\rm min})/2}\,,
\end{equation}
where $A$ is a normalization constant, using
\begin{equation}
\frac{dP}{d\sin^2(\theta)}=\int\!dk\!\int\!\!dz\,\delta(\sin^2(\theta)-f(k,z)) Ae^{-(\chi^2(k,z)-\chi^2_{\rm min})/2}\,,
\end{equation}
where $f(k,z)$ is one of the functions on the righthand side of Eqs.\,(\ref{eq:s2asx-IH}). The results are shown in Figs.\,(\ref{distna}). 

As would be expected, none of the distributions is exactly Gaussian. The largest contribution to the minimum $\chi^2$ is associated with $\sin^2(\theta_a)$ and the influence of terms beyond the quadratic expansion of $\chi^2(k,z)$ can be seen in the shape of this distribution.

\begin{figure}[h]\centering
\includegraphics[width=2.75in]{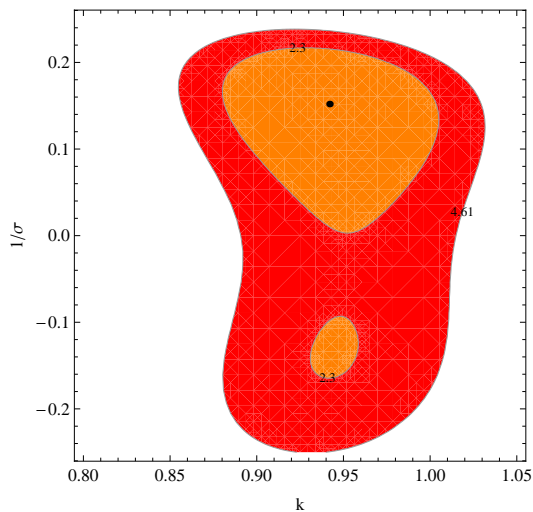}
\caption{The 68.3\% and 90.0\% confidence contours for the fit using Eqs.\,(\ref{s2a}), (\ref{s2x}) and (\ref{s2s}) are shown in red and orange, respectively. The (black) dot indicates the $\chi^2$ minimum.} \label{abd}
\end{figure}

If we use Eqs.\,(\ref{s2a}), (\ref{s2x}) and (\ref{s2s}), the fit to the data has two local minima. The lowest of these gives $\chi^2_{\rm min}=0.506$, $|k|_{\rm min}=0.942$ and $z_{\rm min}=0.152$. At the minimum values of $|k|$ and $z$, $\sin^2(\theta_a)=0.423\,(40.5^\circ)$, $\sin^2(\theta_x)=0.013\, (6.55^\circ)$ and $\sin^2(\theta_s)=0.311\,(33.9^\circ)$. At the other minimum, where $\chi^2=2.73$, $\sin^2\theta_s$ and $\sin^2\theta_x$ are slightly different, but $\sin^2\theta_a=0.567$. This is reflected in the individual mixing angle distributions. The confidence contours for this case are shown in Fig.\,\ref{abd}.

\begin{figure}[h,t]\centering
\includegraphics[width=2.3in]{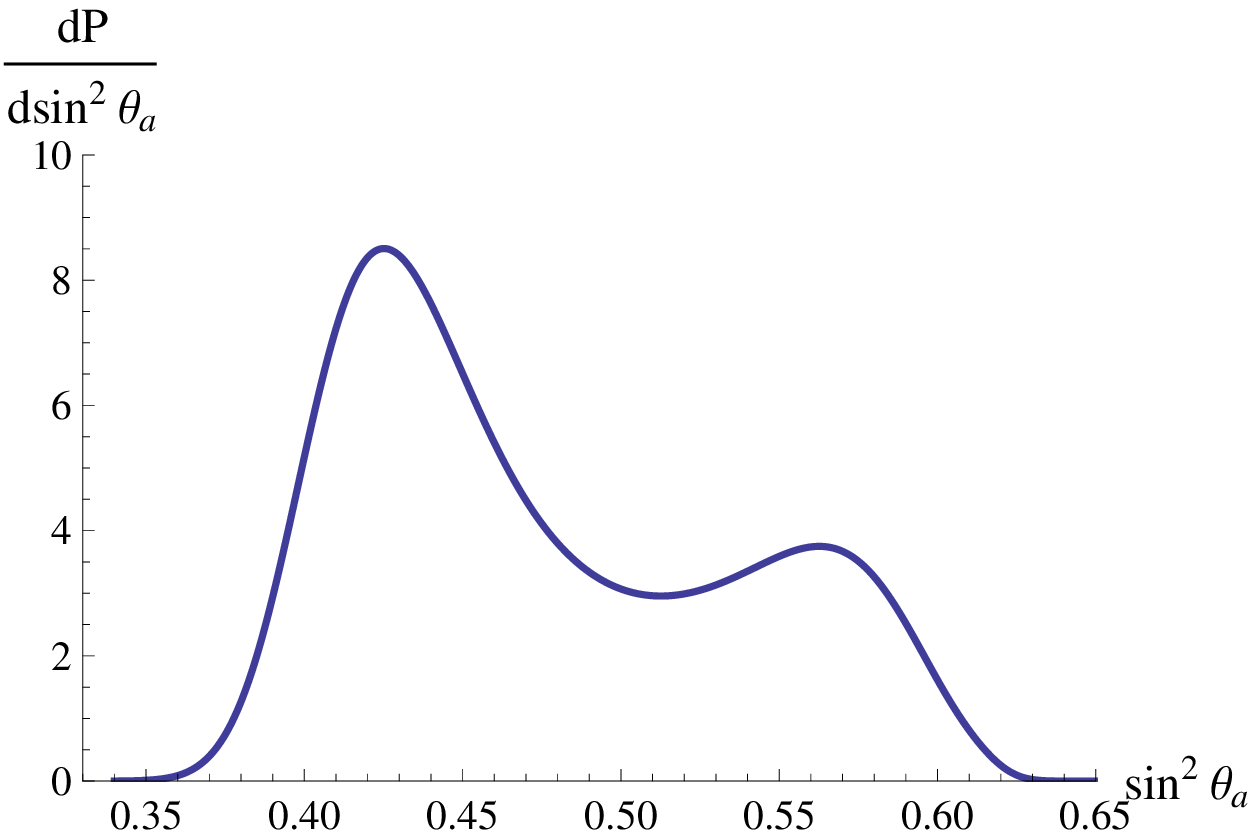}
\hfill\includegraphics[width=2.3in]{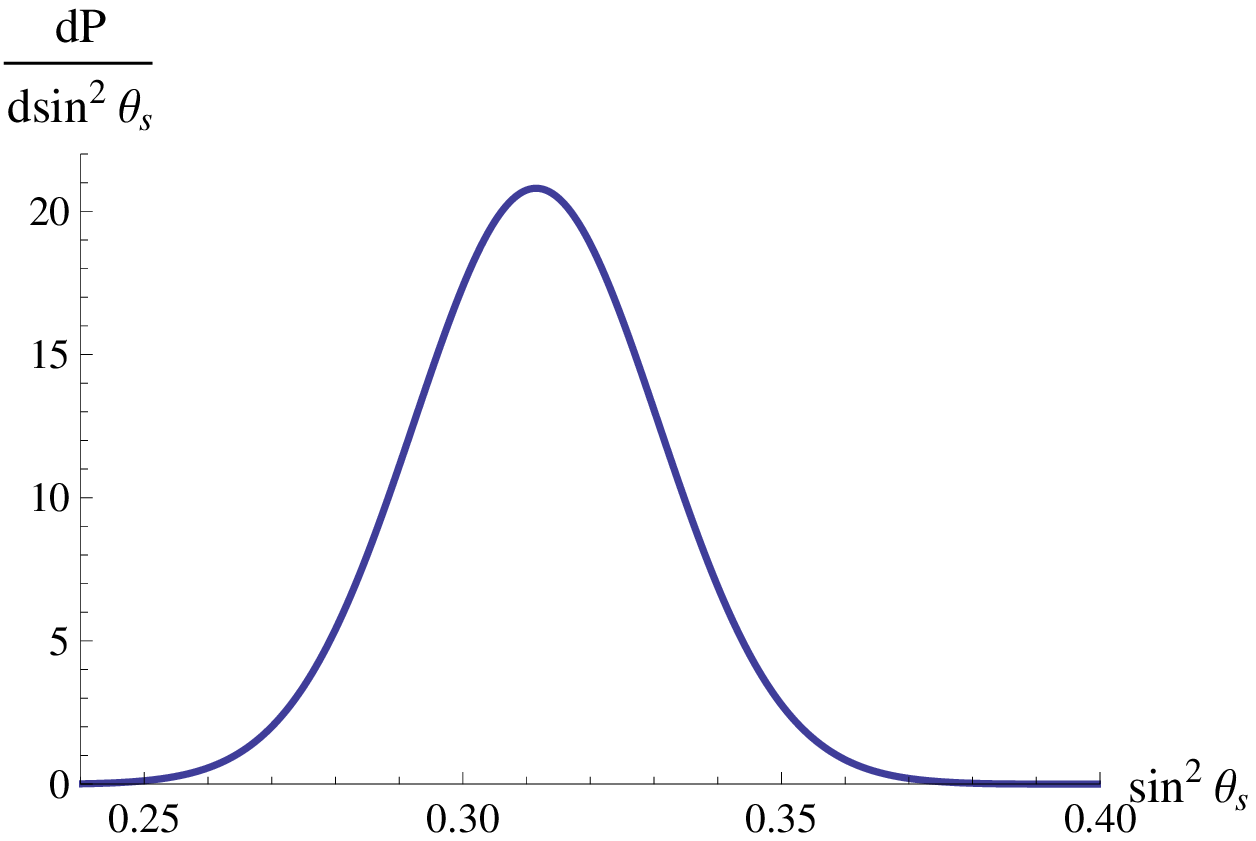}\hfill\vspace{6 pt} 
\hfill\includegraphics[width=2.3in]{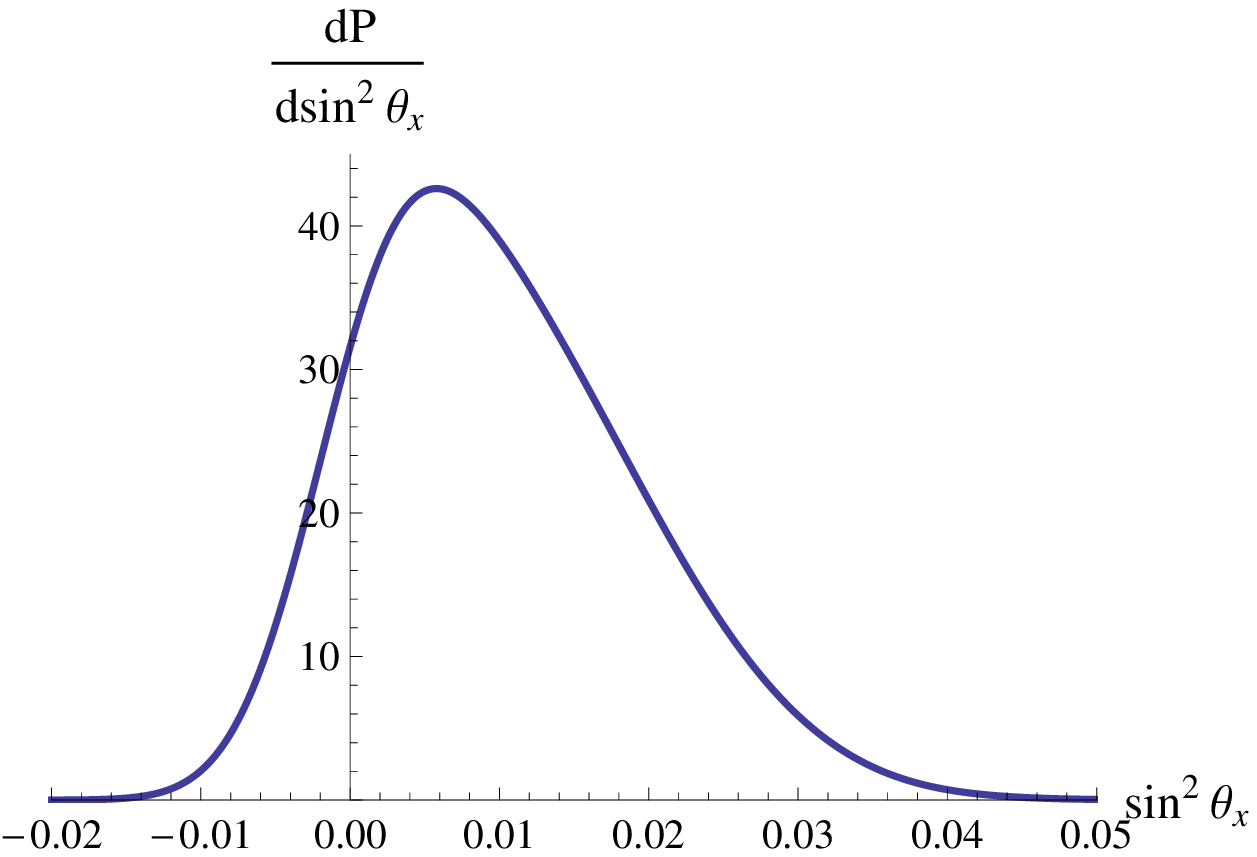}\hfill
\caption{The distributions of the $\sin^2\theta_i$ obtained using Eqs.\,(\ref{s2a}), (\ref{s2x}) and (\ref{s2s}) are shown.}\label{distnb}
\end{figure}

The distributions of this set of mixing angles are shown in Figs.\,(\ref{distnb}). Here, too, the largest contribution to the minimum $\chi^2$ is associated with $\sin^2(\theta_a)$ and the effect of the second local minimum this is reflected in the distortion on the high side of the probability distribution.

Alternately we can fit for $k$ using the values of $\sin^2\theta_a$ and $\sin^2\theta_s$ from Ref.\cite{Fogli-08} but replace $\sin^2\theta_x$ with the value for $\sin^2(2\theta_a)\,\sin^2(2\theta_x)$ published by the MINOS collaboration \cite{MINOS}. They report $\sin^2(2\theta_a)\,\sin^2(2\theta_x)\,\simeq\,0.18\pm0.13$ for inverted hierarchy and, for normal hierarchy, $\simeq\,0.11\pm0.09$. From the inverted hierarchy result we get $|k|=2.10\pm\,0.10$ with a $\chi^2$ of $1.86$ or $|k|=0.94\pm\,0.15$ with a $\chi^2$ of $1.35$.

This was all for inverted hierarchy.  Normal hierarchy, with $m1$ equal to zero, gives, after a lot of work, exactly the results of inverted hierarchy, (\ref{r1}), (\ref{r3p}), (\ref{r3a}). The parameter $\sigma$ is a different function than (\ref{r2}),
\begin{equation}\label{sigmaNH}
\sigma\,=\,\frac{s_a(1+k^2)-c_a}{k(s_a+c_a)}\,,
\end{equation}
but this just amounts to a reparameterization of (\ref{eq:s2asx-IH}) with no physical consequence. Using the MINOS number for normal hierarchy we find the same values, including the errors, for $|k|$ as for the inverted hierarchy MINOS number. The $\chi^2$ values are smaller at $1.42$ or $0.80$.

With either MINOS value and for either value of $|k|$ the fitted value of $\sin^2\theta_s$ is stable at $0.312$, the fitted value of $\sin^2\theta_a$ varies only slightly from $0.46$ for the larger $|k|$ to $0.42$ for the smaller value, but $\sin^2\theta_x$ is less than $0.001$ for the larger $|k|$ but equal to  $0.015$ for the smaller.

\section{Summary}

A hidden $\mathcal{Z}_2$ symmetry, as given by Eq.\,(\ref{GMG}), results in only two possible sets of conditions on the neutrino mixing angles.  Assuming $\theta_s\,<\,\pi/4$ then either
\begin{subequations}
\begin{eqnarray}
s_x\,&=&\,\frac{c_x}{k}(s_a-c_a)\,, \label{sol1}  \\
\tan\theta_s\,&=&\,-\frac{c_x}{k}(s_a+c_a)\,, \label{sol2}
\end{eqnarray}
\end{subequations}
with confidence contours shown in Fig.\,\ref{abc}, or
\begin{subequations}
\begin{eqnarray}
s_x\,&=&\,\frac{c_x}{k}(s_a-c_a)\,, \label{sol3}  \\
\tan\theta_s\,&=&\,\frac{k}{c_x(s_a+c_a)}\,,\label{sol4}
\end{eqnarray}
\end{subequations}
with the confidence contours shown in Fig.\,\ref{abd}.

\section*{Acknowledgments}

SFG was supported by the China Scholarship Council (CSC). DAD and SFG were supported in part by the U. S. Department of Energy under grant No. DE-FG03-93ER40757. WWR was supported in part by the National Science Foundation under Grant PHY-0555544. DAD is a member of the Center for Particles and Fields and the Texas Cosmology Center. It is always a pleasure to thank Sacha Kopp and Karol Lang for helpful discussions regarding the MINOS data. WWR thanks Jim Linnemann for an informative conversation about the procedures used in Sec.\,(3).


\begin{thebibliography}{99}
\bibitem{PMNS}
B.~Pontecorvo,  Sov.\ Phys.\ JETP {\bf 6}, 429 (1957) [Zh.\ Eksp.\ Teor.\ Fiz.\  {\bf 33}, 549 (1957)]; Z.~Maki, M.~Nakagawa and S.~Sakata, Prog.\ Theor.\ Phys.\  {\bf 28}, 870 (1962).
\bibitem{Fogli-08}
G.~L.~Fogli, E.~Lisi, A.~Marrone, A.~Palazzo and A.~M.~Rotunno, Phys.\ Rev.\ Lett.\  {\bf 101}, 141801 (2008) [arXiv:0806.2649 [hep-ph]] and G.~L.~Fogli, E.~Lisi, A.~Marrone, A.~Palazzo and A.~M.~Rotunno,  arXiv:0809.2936 [hep-ph]; G.~L.~Fogli {\it et al.},  Phys.\ Rev.\  D {\bf 78}, 033010 (2008) [arXiv:0805.2517 [hep-ph]] and references therein.
 \bibitem{nu2008}
See, for instance, the experimental reports at XXIII International Conferences on ``Neutrino Physics and Astrophysics'' (Neutrino 2008), Christchurch, New Zealand, May 25-31, 2008. Web link: \url{http://www2.phys.canterbury.ac.nz/~jaa53}
\bibitem{Lam:2006wm} C.~S.~Lam, Phys.\ Rev.\  {\bf D74}, 113004 (2006).
\bibitem{Lam:2008sh} C.~S.~Lam,  Phys.\ Rev.\  D {\bf 78}, 073015 (2008) [arXiv:0809.1185 [hep-ph]].
\bibitem{Grimus:2009pg} W.~Grimus, L.~Lavoura, P.~O.~Ludl, J.\ Phys.\ G {\bf G36}, 115007 (2009).
\bibitem{GHY} S.-~F.~Ge, H.~J.~He and F.~R.~Yin,  JCAP {\bf 1005}, 017 (2010) [arXiv:1001.0940 [hep-ph]].
\bibitem{mu-tau} P.~F.~Harrison and W.~G.~Scott, Phys.\ Lett.\  B {\bf 547}, 219 (2002) [arXiv:hep-ph/0210197], R.~N.~Mohapatra, JHEP {\bf 0410}, 027 (2004) [arXiv:hep-ph/0408187], R.~N.~Mohapatra, S.~Nasri and H.~B.~Yu, Phys.\ Lett.\  B {\bf 615}, 231 (2005)  [arXiv:hep-ph/0502026], T.~Kitabayashi and M.~Yasue, Phys.\ Lett.\  B {\bf 621}, 133 (2005) [arXiv:hep-ph/0504212], R.~N.~Mohapatra and W.~Rodejohann, Phys.\ Rev.\  D {\bf 72}, 053001 (2005) [arXiv:hep-ph/0507312], Z.~Z.~Xing, H.~Zhang and S.~Zhou, Phys.\ Lett.\  B {\bf 641}, 189 (2006) [arXiv:hep-ph/0607091].
\bibitem{MinimalSeesaw} P.~H.~Frampton, S.~L.~Glashow and T.~Yanagida,  Phys.\ Lett.\  B {\bf 548}, 119 (2002) [arXiv:hep-ph/0208157]; M.~Raidal and A.~Strumia,  Phys.\ Lett.\  B {\bf 553}, 72 (2003) [arXiv:hep-ph/0210021].
\bibitem{CPGDR} S.-~F.~Ge, D.~A.~Dicus, and W.~W.~Repko, ``A $\mathbb{Z}_2$ symmetry prediction for the Dirac $CP$ phase of neutrino mixing,'' in preparation.
\bibitem{DGR} These conditions for one vanishing neutrino mass and $k=-1$ were derived in: D.~A.~Dicus, S.-~F.~Ge and W.~W.~Repko,  Phys.\ Rev.\ D{\bf 82}, 033005 (2010) arXiv:1004.3266 [hep-ph].
\bibitem{BDHL} V.~Barger, D.~A.~Dicus, H.~J.~He and T.~J.~Li,  Phys.\ Lett.\  B {\bf 583}, 173 (2004) [arXiv:hep-ph/0310278].
\bibitem{MINOS}P. Adamson {\it et. al.}, Phys. Rev. Lett. {\bf 103}, 261802-1 (2009).
\end{thebibliography}
\end{document}